\documentclass{rspublic}
\usepackage{epsf}
\usepackage{graphicx}
\usepackage{rotating}

\begin{document}

\title[Violence in the Hearts of Galaxies]{Violence in the Hearts of Galaxies - Aberration or
Adolescence?}

\author[C.G. Mundell]{Carole G. Mundell}

\affiliation{Astrophysics Research Institute, Liverpool John Moores
University, Twelve Quays House, Egerton Wharf, Birkenhead, CH41 1LD, UK}

\newcommand{\mn}{{\em Mon. Not. R. Astr. Soc.}}
\newcommand{\ana}{{\em Astron. Astrophys.}}
\newcommand{\apj}{{\em Astrophys. J.}}
\newcommand{\apjs}{{\em Astrophys. J. Suppl.}}
\newcommand{\aj}{{\em Astr. J. }}
\newcommand{\pasp}{{\em Publ. Astron. Soc. Pac. }}
\newcommand{\araa}{{\em A. Rev. Astron. \& Astrophys. }}
\def\gtrsim{\mathrel{\hbox{\rlap{\hbox{\lower4pt\hbox{$\sim$}}}\hbox{$>$}}}}

\label{firstpage}
\maketitle
\begin{abstract}{Black holes;AGN;quasars;galaxy evolution}
Violent activity in the nuclei of galaxies has long been considered a
curiosity in its own right; manifestations of this phenomenon include
distant quasars in the early Universe and comparatively nearby Seyfert
galaxies, both thought to be powered by the release of gravitational
potential energy as material from the host galaxy accretes onto a
central supermassive black hole (SMBH).  Traditionally, the broader
study of the formation, structure and evolution of galaxies has
largely excluded active galactic nuclei. Recently however, this
situation has changed dramatically, both observationally and
theoretically, with the realisation that the growth and influence of
the SMBH, the origin and development of galaxies and nuclear activity
at different epochs in the Universe may be intimately related.  The
most spectacular fireworks seen in distant quasars, may be relatively
easy to explain since the era of greatest quasar activity seems to
coincide with turbulent dynamics at the epoch of galaxy formation in
the young, gas-rich Universe.  Ubiquitous black holes are believed to
be a legacy of this violent birth.  Alternatively, black holes may be
the seeds which drive galaxy formation in the first place. Closer to
home, and hence more recently in the history of the Universe, a
fraction of comparatively ordinary galaxies, similar to our own, have
re-ignited their central engines, albeit at a lower level of
activity. Since these galaxies are more established than their younger
and more distant counterparts, the activity here is all the more
puzzling. Whatever the mechanisms involved, they are likely to play an
important role in galaxy evolution.  I review the intriguing evidence
for causal links between supermassive black holes, nuclear activity
and the formation and evolution of galaxies, and describe opportunities
for testing these relationships using the next generation of
earth-bound and space-borne astronomical facilities.
\end{abstract}

\section{Introduction}

Over the last 50 years, astronomers have been intrigued by enormously
energetic objects called Active Galactic Nuclei (AGN), a violent
phenomenon occurring in the nuclei, or central regions, of some
galaxies with intensities and durations which cannot easily be
explained by stars, thus providing some of the first circumstantial
evidence for theoretically-predicted supermassive black holes. Despite
their intriguing properties they were largely viewed as interesting
but unimportant freaks in the broader study of galaxy formation and
evolution, leading astronomers studying the properties of galaxies to
exclude the small fraction of galaxies with active centres as
irritating aberrations.  Here I describe the discovery of AGN and the
variety of classifications that followed; I describe some features of
unifying models of the central engine that attempt to explain the
varied properties of different AGN classes that give rise to the
classification.  The search for supermassive black holes in AGN and
non-active galaxies is discussed along with the developing realisation
that all galaxies with significant bulge components might harbour
dormant supermassive black holes as remnants of a past adolescent
period of quasar activity and therefore posses the potential to be
re-triggered into activity under the right conditions, making nuclear
activity an integral part of galaxy formation and evolution.

\section{The Early Studies of Active Galactic Nuclei}

The discovery of AGN began with the development of radio astronomy
after World War II when hundreds of sources of radio waves on the sky
were detected and catalogued (e.g. Third Cambridge Catalogue (3C) -
Edge {\em et al.} 1959 and its revision (3CR) - Bennett 1961), but the
nature of these strong radio emitters was unknown.  Astronomers at
Palomar attempted to optically identify some of the catalogued radio
sources; Baum \& Minkowski (1960) discovered optical emission from a
faint galaxy at the position of the radio source 3C295 and, on
studying the galaxy's spectrum, or cosmic bar-code, measured its
redshift and inferred a distance of 5000 million light years, making
it the most distance object known at that time.  Distances can be
inferred from Hubble's law whereby the more distant an object, the
faster it appears to be receding from us, due to the expansion of the
Universe.  Chemical elements present in these objects emit or absorb
radiation at known characteristic frequencies and when observed in a
receding object, the observed frequency is reduced or {\em redshifted}
due to the Doppler effect; the same physical process that causes a
receding ambulance siren to be lowered in pitch after it passes the
observer.

Attempts to find visible galaxies associated with other strong radio
sources such as 3C48, 3C196 and 3C286 failed and only a faint blue,
star-like object at the position of each radio source was found - thus
leading to their name `quasi-stellar radio sources', or `quasars' for
short. The spectrum of these quasars resembled nothing that had
previously been seen for stars in our Galaxy and these blue points
remained a mystery until Maarten Schmidt (1963) concentrated on 3C273,
for which an accurate radio position was known (Hazard, Mackey \&
Shimmins 1963). The optical spectrum of the blue source associated
with the radio emitter seemed unidentifiable until Schmidt realised
that the spectrum could be clearly identified with spectral lines
emitted from hydrogen, oxygen and magnesium atoms if a redshift
corresponding to 16\% of the speed of light was applied.  The same
technique was applied successfully to 3C48 (Greenstein \& Matthews
1963) and demonstrated that these objects are not members of our own
galaxy but lie at vast distances and are super-luminous. Indeed, the
radiation emitted from a quasar (L $\gtrsim$10$^{13}$~L$_{\odot}$, where
the Sun's luminosity is L$_{\odot}$ = 3.8~$\times$~10$^{26}$ Watt) is
bright enough to outshine all the stars in its host galaxy. Such
energies cannot be produced by stars alone and it was quickly realised
that the release of gravitational potential energy from material
falling towards, or being accreted by, a {\em supermassive black hole}
at the galaxy centre, $\sim$100 times more energy efficient than
nuclear fusion, was the only effective way to power such prodigious
outputs (Lynden-Bell 1969).

A black hole is a region of space inside which the pull of gravity is
so strong that nothing can escape, not even light.  Two main kinds of
black holes are thought to exist in the Universe. {\em Stellar-mass}
black holes arise from the the collapsed innards of a massive star
after its violent death when it blows off its outer layers in a
spectacular supernova explosion; these black holes have mass slightly
greater than the Sun but are compressed into a region only a few
kilometres across. In contrast, {\em supermassive black holes}, which
lurk at the centres of galaxies, are 10 million to 1000 million times
more massive than the Sun and contained in a region about the size of
the Solar System.  The emission of radiation from a supermassive black
hole appears at first to be contradictory; however, the energy
generating processes take place outside the black hole's
point-of-no-return, or {\em event horizon}. The mechanism involved is
the conversion of gravitational potential energy into heat and light
by frictional forces within a disk of accreting material, which forms
from infalling matter that still possesses some orbital energy, or
angular momentum, and so cannot fall directly into the black hole.

Radiation from AGN is detected across the electromagnetic spectrum and
today, nuclear activity in galaxies has been detected over a wide
range of luminosities, from the most distant and energetic quasars, to
the weaker AGN seen in nearby galaxies, such as Seyferts (Seyfert
1943), and even the nucleus of our own Milky Way.

\section{AGN Orientation - Looking at it from All Angles}

\begin{figure}
\epsfclipon
\epsfxsize=\textwidth
\epsffile{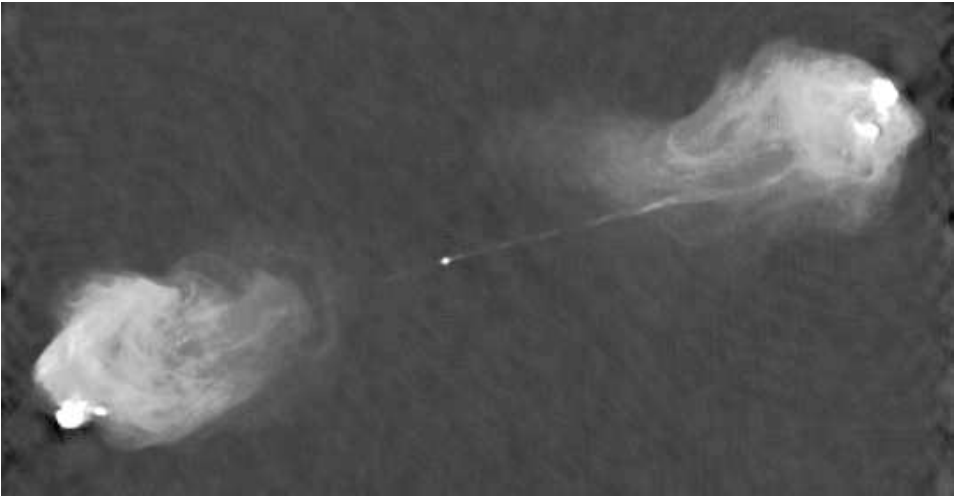}
\caption{6-cm radio image of the classic radio galaxy Cygnus A (courtesy Chris Carilli)}   
\label{cygA}
\end{figure}

After the initial discovery of radio-loud AGN, the advent of radio
interferometry soon led to detailed images of these strong radio
emitters (e.g., Bridle \& Perley 1984; Bridle {\em et al.} 1994) which
revealed remarkable long thin jets of plasma emanating from a central
compact nucleus and feeding extended lobes, often at considerable
distances from the AGN, millions of light years in the most extreme
cases.  The radio emission is synchrotron radiation produced by
electrons spiraling around magnetic fields in the ejected plasma;
figure
\ref{cygA} shows a radio image of the classic radio galaxy Cygnus A in
which the nucleus, jets and lobes are visible. These dramatic jets and
clouds of radio-emitting plasma were interpreted as exhaust material
from the powerful central engine (Scheuer 1974; Blandford \& Rees
1974).

\subsection{Too fast to believe - the remarkable jets in radio-loud AGN}

The sharpest radio images, made repeatedly over many years using
networks of radio telescopes spanning the globe, resulted in `movies'
of the motion of material in the jets. The blobs of plasma in these
jets were apparently being ejected at many times the speed of light,
{\em c}, appearing to violate fundamental laws of physics. It was
quickly realised that such {\em superluminal} motion, was an optical
illusion caused by the plasma moving at {\em relativistic} speeds,
i.e. $\gtrsim$~0.7{\em c}, and being ejected towards us at an angle
close to our line of sight (e.g., Blandford \& Rees 1978).
Relativistic motion appears to be present for jet matter over hundreds
of thousands of light years and the detailed physical driving
mechanisms remain an area of active study.  The relativistic motion of
jet matter has an enormous impact on the appearance of these objects
and is possibly the single-most important contributor to the variety
of observed morphological types.

   \begin{figure}
\epsfclipon
\epsfxsize=\textwidth
\epsffile{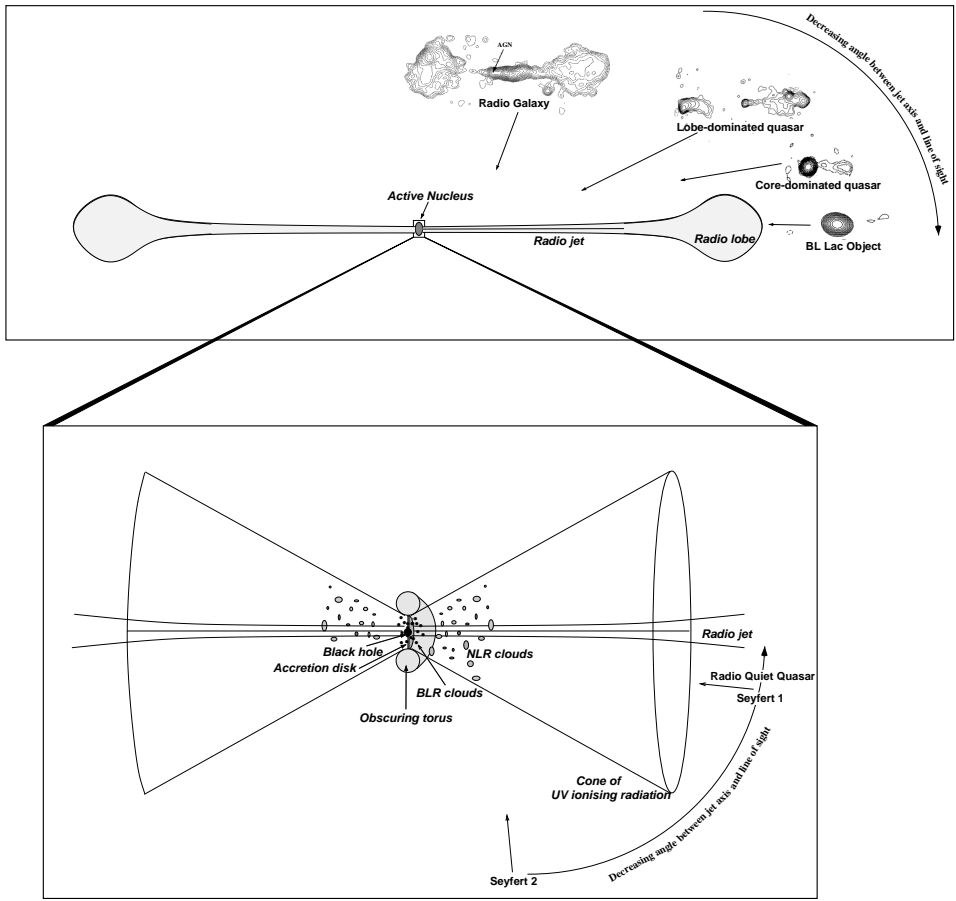}
\caption{Top: Radio-loud unification scheme, in which the observed AGN 
type depends on the observer's viewing angle to the ejection axis of
the radio jet. Bottom: Sketch of an AGN central engine with a central
black hole surrounded by (a) an accretion disc that emits cones of
ultraviolet ionising radiation and defines the radio jet launch
direction, (b) a torus of dust a gas that accounts for the different
observed kinds of radio-quiet AGN by blocking our view of the
accretion disc and dense, rapidly-moving ionised gas clouds in the
broad-line region (BLR) when viewed edge-on (type 2
objects). Less-dense ionised clouds in the narrow-line region (NLR)
lie above the plane of the torus and are visible from all angles.}
\label{orient}
\end{figure}

The fast motion of jet material also causes extreme apparent
brightening, or Doppler boosting, of the radiation and greatly
amplifies any flickering, or variability, in the light levels.  Today,
the wide range of observed radio structures, brightnesses and levels
of variability can be understood in terms of the angle at which we
view the high-speed plasma jet. {\em Radio galaxies} like Cygnus A are
orientated perpendicular to our line of sight, lying in the plane of
the sky, appear rather symmetrical, and as expected, show no
variability or superluminal motion. At the other extreme are bright,
compact and highly variable {\em BL Lac objects}, which are being
observed head-on. Figure \ref{orient} shows a sketch of this model in
which a jet of plasma is ejected from either side of the central
engine at relativistic speeds; object classification depends on the
angle of the jet to our line of sight.  Objects viewed at intermediate
angles are seen as either extended, `lobe-dominated' quasars or
relatively compact, `core-dominated' quasars (see also Urry
\& Padovani 1995).

\subsection{Obscuring-Doughnuts in Radio-Quiet AGN}

{\em Radio-quiet} quasars and Seyferts are known to be $\sim$10 times
more common, but 100 to 1000 times weaker at radio wavelengths and
significantly less extended than their {\em radio-loud} cousins
(Goldschmidt {\em et al.}  1999), but orientation still has important
effects, this time on the optical properties.  Optical spectroscopy
provides a powerful diagnostic tool for the physical conditions in
astronomical objects; as described earlier chemical elements have a
characteristic spectral signature and physical conditions within a gas
can be inferred from distortions of this chemical bar-code. In
particular, broadening of the spectral lines indicates a spread in
gas-cloud velocities, whilst the relative brightnesses of spectral
lines indicate the intensity of ultraviolet radiation incident upon
the gas.

Measurements of the optical spectra of Seyfert nuclei show spectral
lines from gas ionised (i.e. gas in which atoms have been stripped of
one or more electrons) by strong ultraviolet radiation that is too
intense to be produced by a collection of stars and is instead thought
to originate from the accretion disk.  All Seyfert nuclei contain a
region of ionised gas, the Narrow Line Region (NLR), extending over
several hundred light years where the spectral line-widths correspond
to gas velocties of a few hundred km~s$^{-1}$ and densities are
moderate (electrons per unit volume
$n_e$$\sim$10$^3$$-$10$^6$~cm$^{-3}$). Closer in, within $\sim$0.1
light year of the black hole, is the Broad-Line Region (BLR), a much
denser region of gas ($n_e$$\sim$10$^9$~cm$^{-3}$) that shows gas
velocities up to 10,000~km~s$^{-1}$. Seyferts were originally
classified into two types; type-1 Seyferts that show evidence for both
a BLR and an NLR, and type-2 Seyferts that show only an NLR
(Khachikian \& Weedman 1971, 1974).

\begin{figure}
\epsfclipon
\epsfxsize=12.6cm
\epsffile{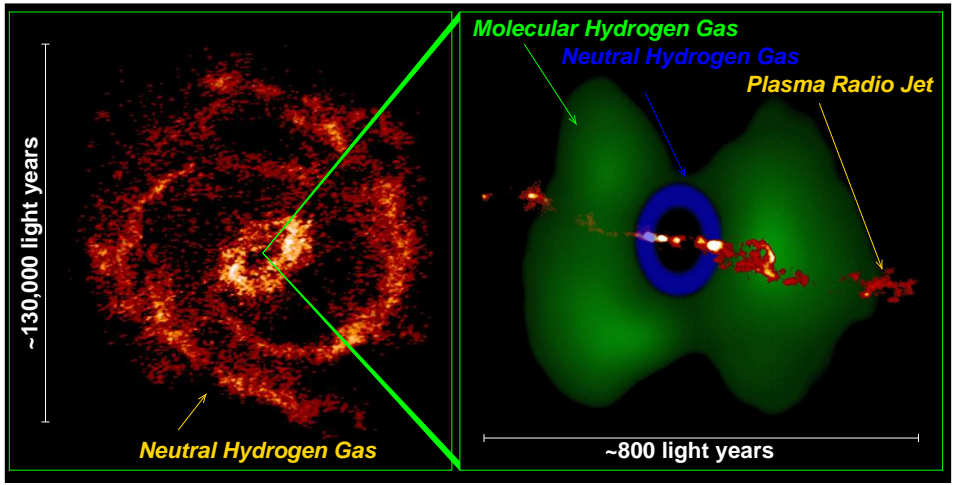}
   \caption{Left: Radio image of neutral hydrogen gas in the spiral
   Seyfert host galaxy NGC~4151 (Mundell {\em et al.} 1999); Right:
   composite image of the central regions of NGC~4151 showing a
   1.4-GHz radio image of the well-collimated plasma jet surrounded by
   an obscuring torus of molecular hydrogen imaged at 2.2$\mu$m
   (Fernandez et al. 1999) and an inferred inner ring of neutral
   hydrogen from absorption measurements (Mundell {\em et al.}  2002).}
\label{myseyfert}
   \end{figure}

The mystery of the missing BLRs in type-2 Seyferts was solved
elegantly in 1985 when Antonucci \& Miller discovered a hidden BLR in
the {\em scattered} light spectrum of the archetypal Seyfert 2 galaxy
NGC~1068, which closely resembled that of a Seyfert type 1. This
discovery led to the idea that the BLR exists in all Seyferts and is
located inside a doughnut, or torus, of molecular gas and dust; our
viewing angle with respect to the torus then explains the observed
differences between the unobscured, broad-line Seyfert 1s, viewed
pole-on, and the obscured, narrow-line Seyfert 2s, viewed
edge-on. Hidden Seyfert 1 nuclei can then be seen in reflected light
as light photons are scattered into the line of sight by particles
above and below the torus acting like a ``dentist's mirror''
(Antonucci \& Miller, 1985; Tran 1995; Antonucci 1993; Wills
1999). The lower panel of Figure
\ref{orient} shows a  sketch of a Seyfert nucleus with the
different types of AGN observed as angle between line of sight and
torus axis increases. Figure \ref{myseyfert} shows an image of the
molecular torus in NGC~4151, surrounding the mini, quasar-like radio
jet emanating from the centre of the galaxy, as predicted by the
unification scheme.

Radio quiet quasars also have broad and narrow lines and are
considered to be the high luminosity equivalents of Seyfert type 1
galaxies. A population of narrow-line quasars, high luminosity
equivalents to obscured Seyfert 2s, are predicted by the unification
scheme but, until now, have remained elusive. New optical and infrared
sky surveys are beginning to reveal a previously undetected population
of red AGN (Cutri {\em et al.}  2001) with quasar type 2 spectra
(Djorgovski {\em et al.}  1999) and weak radio emission (Ulvestad {\em
et al.}  2000). A significant population of highly obscured but
intrinsically luminous AGN would alter measures of AGN evolution, the
ionisation state of the Universe and might contribute substantially to
the diffuse infrared and X-ray backgrounds.

\subsection{Further unification?}

The presence of gas emitting broad and narrow optical lines in
radio-loud AGN and the discorvery of mini radio jets in Seyferts
(e.g. Wilson \& Ulvestad 1982)  led to further consistency between
the two unification schemes. Nevertheless, the complete unification of
radio-loud and radio-quiet objects remains problematic, particularly
in explaining the vast range in radio power and jet extents, and might
ultimately involve the combination of black hole properties, such as
accretion rate, black hole mass and spin, and orientation (Wilson \&
Colbert, 1995; Boroson 2002).

 \begin{figure}
\epsfclipon
\epsfxsize=\textwidth
\epsffile{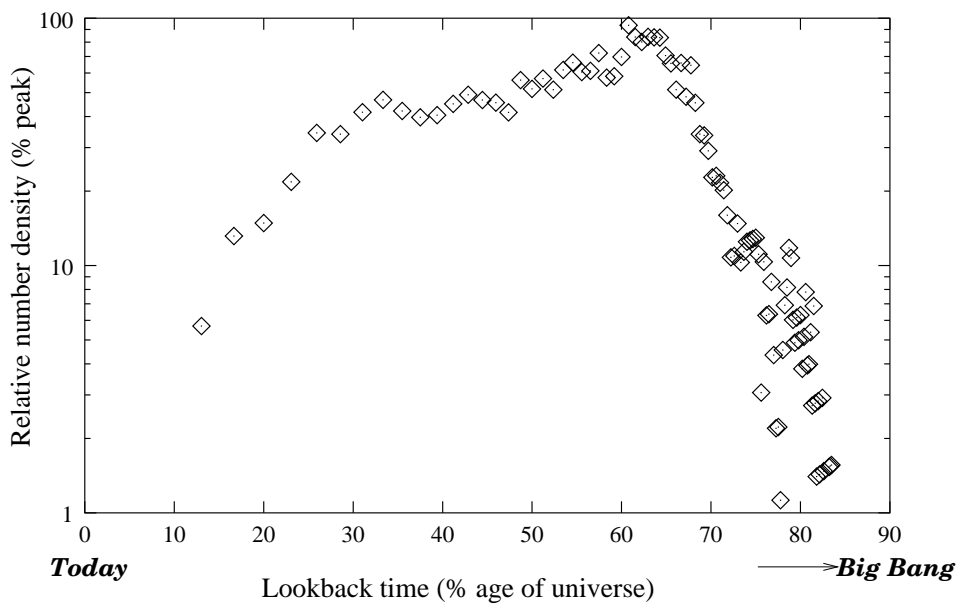}
   \caption{Relative number of galaxies per unit volume of space (as a
   fraction of the peak value) detected in the SLOAN Digital Sky
   Survey, as a function of {\em look-back time} i.e. time running
   backwards from now to the Big Bang [using data from Schneider {\em
   et al.}  (2002)].}
\label{lookback}
   \end{figure}

\section{Searching for Supermassive Black Holes}

Although incontrovertible observational proof of the existence of
supermassive black holes (SMBHs) has yet not been found, evidence is
mounting to suggest the presence of massive dark objects, or large
mass concentrations at the centres of galaxies.  Black holes, by
definition, cannot be `seen' and instead one must look for the
consequences of their presence.  The presence of SMBHs has been
inferred indirectly from the energetics of accretion required to power
luminous AGN and explain rapid flux variability and, more directly,
from kinematic studies of the influence of the black hole's
gravitational pull on stars and gas orbiting close to it in the
central regions of both active and non-active galaxies.  Theoretical
models rule out alternatives to a supermassive black holes such as
collections of brown or white dwarf stars, neutron stars or
stellar-mass black holes which would merge and shine or evaporate too
quickly (Maoz 1995, 1998; Genzel {\em et al.} 1997, 2000).

\subsection{Quasar lifetimes and the black hole legacy}

Soon after the discovery of quasars it became clear that they were
most common when the Universe was relatively young with the peak of
the quasar epoch at redshift z$\sim$2.5 or a {\em look-back time} of
65\% of the age of the Universe (See Figure \ref{lookback}); today
bright quasars are rare and weaker Seyferts dominate instead.  The
number of dead quasars or relic, dormant black holes left today can
estimated by applying some simple arguments to the quasar
observations. Soltan (1982) integrated the observed light emitted by
quasars, and, assuming the power source for quasar light is accretion
of material by a supermassive black hole with a mass-to-energy
conversion efficiency of 10\% and that the black hole grows during the
active phase, predicted the total mass in relic black holes
today. Knowing the number of galaxies per unit volume of space
(e.g. Loveday {\em et al.} 1992), if one assumes that all galaxies
went through a quasar phase at some time in their lives, then each
galaxy should, on average, contain a $\sim$10$^8$ M$_{\odot}$ black
hole as a legacy of this violent, but short-lived period ($\sim$10$^7$
to $\sim$10$^8$ years). Alternatively, if only a small fraction of
galaxies went through a quasar phase, the active phase would have
lasted lasted longer ($>$10$^9$ years) and the remnant SMBHs would be
relatively rare, but unacceptably massive ($>$10$^{9}$~M$_{\odot}$)
(e.g. Cavaliere et al., 1983; Cavaliere \& Szalay 1986, Cavaliere \&
Padovani 1988).

More complex models including quasar evolution (e.g. Tremaine 1996;
Faber {\em et al.} 1997) and the effects of galaxy growth (e.g. Haehnelt
\& Rees 1993) favour short-lived periods of activity in many
generations of quasars, or a mixture of continuous and recurrent
activity (Small \& Blandford 1992; Cen 2000; Choi, Yang \& Yi 2001).
The complex physics of accretion and black hole growth, however,
remain an area of active study (e.g. Blandford \& Begelman 1999;
Fabian 1999).  Nevertheless, the range of black hole mass of interest
is thought to be M$_{\bullet}$~$\sim$10$^6$ to 10$^{9.5}$M$_{\odot}$,
with the lower mass holes being ubiquitous (Kormendy \& Gebhardt 2001).

\subsection{Irresistible black holes - dynamics of gas and stars}

Although the prodigious energy outputs from powerful quasars offer
strong circumstantial evidence that supermassive black holes exist,
most notably in driving the ejection and acceleration of long,
powerful jets of plasma close to the speed of light (Rees {\em et al.}
1982), it has not, until recently, been possible to make more direct
kinematic measurements of the black hole's gravitational influence.
The mass of a central object, the circular velocity of an orbiting
star and the radius of the orbit are related by Newton's Laws of
motion and gravity. Precise measurements of the velocities of stars
and gas close to the centre of a galaxy are then used to determine the
mass of the central object.

The strongest dynamical evidence for black holes comes from studies of
centre of our own Galaxy and a nearby Seyfert, NGC~4258; a decade of
painstaking observations of a cluster of stars orbiting around the
mildly active centre of the Milky Way, within a radius of 0.07 light
years of the central radio source Sgr A*, suggest a central mass of
M$_{\bullet}$=(2.6$\pm$0.2)$\times$10$^6$ M$_{\odot}$ (Eckart
\& Genzel 1997; Genzel 1997, 2000; Ghez 2000). Discovery of strong
radio spectral lines, or megamasers, emitted from water molecules in a
rapidly rotating nuclear gas disc at the centre of NGC~4258 implies a
centre mass M$_{\bullet}$=(4$\pm$0.1)$\times$10$^7$ M$_{\odot}$
concentrated in a region smaller than 0.7 light years (Miyoshi {\em et
al.} 1995), again small enough to rule out anything other than a black
hole (Maoz 1995, 1998).  Precision measurements of black hole masses
in other galaxies using a variety of techniques, although challenging
and still model dependent, have become increasingly common
(e.g. Maggorian {\em et al.} 1998; Bower {\em et al.} 1998; Gebhardt
{\em et al.}  2000) and now more than 60 active and non-active galaxies
have black hole estimates. 

\section{Black Hole Demographics - the Host Galaxy Connection}

 \begin{figure}
\epsfclipon
\epsfxsize=\textwidth
\epsffile{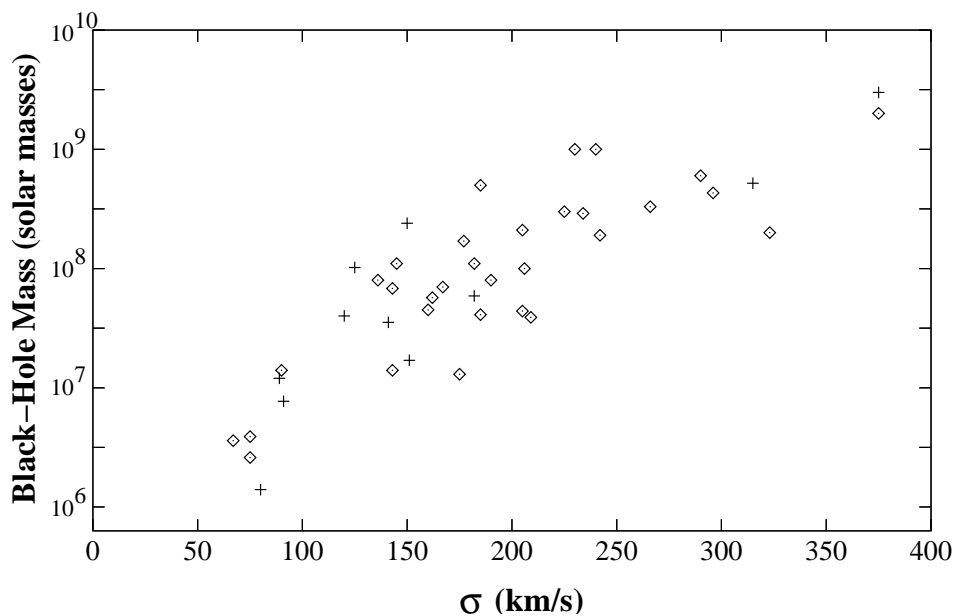}
   \caption{Black hole mass against host galaxy bulge stellar velocity
   dispersion for nearby AGN (open diamonds) and non-active galaxies
   ($+$ symbols) [using data from Kormendy \& Gebhardt (2001) and
   Ferrarese {\em et al.} (2001)].}
\label{msig}
   \end{figure}

In general, galaxies consist of two main visible components - a
central ellipsoidal bulge and a flat disc structure commonly
containing spiral arms - together making a structure resembling two
fried eggs back-to-back.  Elliptical galaxies have no discs and are
dominated by their bulges, maintaining their shapes by the random
motions of their stars; spiral galaxies, like our own Galaxy and
nearby Andromeda have prominent discs and are supported mainly by
rotation, with rotation speeds between 200 km~s$^{-1}$ and 300
km~s$^{-1}$. Some spiral galaxies contain a bar-like structure that
crosses the nucleus; the spiral arms then begin at the ends of the bar
and wind outwards. If the bar is narrow and straight it is classed as
a `strong' bar and if oval-shaped (essentially an elongated bulge) it
is `weak'. Dynamical simulations have revealed that in the region of
the bar, stars do not travel on circular orbits as they do in the
disk, but instead follow more elongated elliptical, or `non-circular'
paths.

With the great progress made recently in measuring the mass of central
supermassive black holes in a significant number of active and
non-active galaxies, correlations with their host galaxy properties
are now possible. Maggorian {\em et al.} (1998) confirmed the
correlation between the brightness of a galaxy bulge (and hence
stellar mass) and the mass of its central black hole (e.g. Kormendy \&
Richstone 1995) establishing a best fit to the linear relation of
M$_{\bullet}$=0.006M$_{bulge}$, despite a large scatter. A much
tighter correlation was subsequently discovered between the velocity
dispersion ($\sigma$) of stars in the host galaxy bulge and the
central black hole mass (e.g. Gebhardt {\em et al} 2000; Ferrarese \&
Merrit 2000).  The velocity dispersion is a measure of the range of
random speeds present in star motions and is potentially a more
reliable galaxy mass indicator than total starlight; the greater the
spread in speeds, the more massive the galaxy bulge. The tightness of the
correlation points to a connection between the formation mechanism of
the galaxy bulge and central black hole although the physics involved
are not yet known.  The M$_{\bullet}$$-$$\sigma$ relation for a
mixture of nearby active and non-active galaxies (Figure \ref{msig}),
measured using a variety of techniques, shows the relationship between
bulge and black hole is very similar for both, although investigations
continue to establish the precise form of the correlation and whether
it is universal for active and non-active galaxies. If universal, this
relationship would provide exciting confirmation that non-active
galaxies contain dormant versions of the same kind of black holes that
power AGN.

No correlation exists between galaxy disc properties and black hole
mass, and disc galaxies without bulges do not appear to contain
supermassive black holes (e.g. Gebhardt {\em et al.} 2001), suggesting
discs form later and are not involved in the process that intimately
links the black hole and bulge.

\section{AGN and their Environment}
\subsection{The violent early Universe}

The relationships between black holes and their host galaxies are
increasingly compelling but unanswered questions remain concerning the
relationship between star formation, galaxy formation, quasar activity
and black hole creation in the early Universe.  Observations of faint
galaxies in the Hubble Deep Field suggested a peak in star formation
history that matches that of the quasar epoch (e.g. Madau {\em et al.}
1996) implying a close link between star formation and quasar
activity.  More recent measurements, however, suggest that the star
formation activity may be constant for redshifts greater than 1 with
the onset of substantial star formation occurring at even earlier
epochs, at redshifts beyond 4.5 (Steidel {\em et al.}  1999). An
increasing number of new quasars are also being found at redshifts
greater than~4 (Fan {\em et al.}  2001; Schneider {\em et al.} 2002)
providing constraints for cosmological models of galaxy formation and
continuing the debate on the relationship between quasar activity,
star formation and the creation of the first black holes (e.g. Haiman
\& Loeb 2001).

The life cycle of an AGN involves a mechanism to trigger the infall of
gas to create an accretion disc and continued fuelling, or
replenishment, of this brightly-shining accretion disc.  A number of
models have suggested that at intermediate to high redshifts it may be
moderately easy to trigger and fuel AGN, where galaxies might be more
gas rich, star formation is vigorous and collisions between galaxies
are common (Haehnelt \& Rees 1993). Kauffman \& Haehnelt (2000)
suggest a model in which galaxy and quasar evolution at early times
was driven by mergers of gas-rich disc galaxies, which drove the
formation and fuelling of black holes and created today's elliptical
galaxies, thereby tying together host galaxy and black hole
properties. As the Universe ages, a decreasing galaxy merger rate and
available gas supply and increasing accretion timescales produce the
decline in bright quasars.

An alternative hypothesis, linking black hole and bulge growth with
quasar activity, involves strong bars in early galaxies (Sellwood
1999); early disc galaxies developed strong bars which were highly
efficient at removing angular momentum from disc gas and funnelling it
towards the centre to feed and grow a black hole. This represents the
bright quasar phase in which the black hole grows rapidly, but on
reaching only a few percent of the mass of the host disc, the central
mass concentration soon destroys the bar due to an increasing number
of stars that follow random and chaotic paths, thereby choking off the
fuel supply and quenching the quasar. In addition, the increase in
random motion in the disc leads to the creation of a bulge. A disc
might be re-built some time later if the galaxy receives a new supply
of cold gas, perhaps from a `minor merger' whereby a small gaseous
galaxy or gas-cloud falls into the main disc and is consumed by the
disc without causing significant disruption, and without significantly
affecting the black hole mass. This scenario nicely accounts for the
relationship between black hole masses and bulge properties and lack
of correlation with disc properties.

An important unknown parameter in these models is the amount of cold
gas in progenitor disc galaxies and how it evolves with time; it is
expected that the Universe was more gas-rich in the past (Barger {\em
et al.}  2001), but observations of neutral hydrogen (H{\sc i}) and
molecular gas such as carbon monoxide (CO) with new generation
facilities, such as Atacama Large Millimeter Array (ALMA), the Giant
Metrewave Radio Telescope (GMRT), the Extended Very Large Array (EVLA)
and the proposed Square Kilometre Array (SKA), will offer exciting
opportunities to measure the gaseous properties of distant galaxies
directly to further our understanding of galaxy formation and
evolution and its relationship to quasar and star-formation activity.

\subsection{Re-activating dormant black holes in nearby galaxies}

While the most luminous AGN might coincide with violent dynamics in
the gas-rich universe at the epoch of galaxy formation (Haehnelt \&
Rees 1993), nuclear activity in nearby galaxies is more problematic
since major galaxy mergers, the collision of two equal-mass disc
galaxies, are less common and galaxy discs are well established;
reactivation of ubiquitous `old' black holes is therefore likely to
dominate.  Host-galaxy gas represents a reservoir of potential fuel
and, given the ubiquity of supermassive black holes, the degree of
nuclear activity exhibited by a galaxy must be related to the nature
of the fuelling rather than the presence of a black hole (e.g.,
Shlosman \& Noguchi 1993; Sellwood \& Moore 1999). Gravitational, or
tidal forces exterted when two galaxies pass close to one another may
play a role in this process, either directly, when gas from the
companion, or outer regions of the host galaxy, is tidally removed and
deposited onto the nucleus, or by causing disturbances to stars
orbiting in the disc and leading to the growth of structures such as
bars, in which stars travel on elliptical paths and drive inflows of
galactic gas (e.g. Toomre \& Toomre 1972; Simkin, Su \& Schwarz 1980;
Shlosman, Frank \& Begelman 1989; Mundell {\em et al.}  1995;
Athanassoula 1992; Mundell \& Shone 1999).

Numerous optical and IR surveys of Seyfert hosts have been conducted
but as yet show no conclusive links between nuclear activity and host
galaxy environment. Neutral hydrogen (H{\sc i}) is an important tracer of
galactic structure and dynamics and may be a better probe of
environment than the stellar component.  H{\sc i} is often the most
spatially extended component of a galaxy's disc so is easily disrupted
by passing companions, making it a sensitive tracer of tidal
disruption (e.g. Mundell et al., 1995). In addition, because gas can
dissipate energy and momentum through shock waves (Mundell \& Shone,
1999), whereas collisions between stars are rare, the observable
consequences of perturbating the H{\sc i} in galactic bars are easily
detectable. However, despite the diagnostic power of H{\sc i}, until
recently few detailed studies of H{\sc i} in Seyferts have been
performed (Brinks \& Mundell 1996; Mundell 1999).

The strength of a galaxy collision, which depends on initial galaxy
properties such as mass, concentration, distance and direction of
closest approach, ranges from the most violent mergers between equal
mass, gas-rich disc galaxies, to the weakest interaction in which a
low mass companion, perhaps on a fly-by path, interacts with a massive
primary. In this minor-merger case the primary disc is perturbed but
not significantly disrupted or destroyed.  Indeed, Seyfert nuclei are
rare in strongly interacting systems, late-type spirals and elliptical
galaxies (Keel et al. 1995; Bushouse 1996) and sometimes show
surprisingly undisturbed galactic discs despite the presence of H{\sc
i} tidal features (Mundell et al. 1995).  Seyfert activity may
therefore involve weaker interactions or minor mergers between a
primary galaxy and a smaller companion or satellite galaxy, rather
than violent major mergers (e.g. De Robertis, Yee \& Hayhoe 1998).  A
key question is whether the gaseous properties of normal galaxies
differ from those with Seyfert nuclei and a deep, systematic H{\sc i}
imaging survey of a sample of Seyfert and normal galaxies
is now required.

\section{Unanswered Questions and Prospects for the Future}

Studies of galaxies and AGN are being revolutionised by impressive new
sky surveys, such as SLOAN and 2DF, which have already significantly
increased the number of known galaxies and quasars in the Universe.
In the next decade and beyond, prospects for understanding AGN and
their role in galaxy formation and evolution are extremely promising
given the number of planned new instruments spanning the
electromagnetic spectrum.  

\noindent
$\bullet$We do not yet know whether galaxies grow
black holes or are seeded by them; NGST (Next Generation Space
Telescope) will find the smallest black holes at the earliest times
and allow us to relate them to the first galaxies and stars.

\noindent
$\bullet$ The amount of cold gas in galaxies through cosmic history is
a key ingredient in star-formation, quasar activity and galaxy
evolution models but is still unknown. The study of gas at high
redshifts with ALMA, the GMRT and the EVLA will revolutionise our
understanding its role in these important phenomena and provide
powerful constraints for cosmological models.

\noindent
$\bullet$Current models of AGN physics - fuelling, accretion discs and
the acceleration of powerful radio jets - remain speculative; detailed
studies of X-ray emitting gas, e.g with the highly ambitious X-ray
space interferometer MAXIM, might offer valuable new insight into the
energetics and physical structure of this extreme region.

\noindent
$\bullet$Finally, the detection and detailed study of gravitational
waves, using the space-based detector LISA, from massive black holes
living in black-hole binary systems or in the very act of merging will
prove the existence of SMBHs and perhaps provide insight into the
origin of the difference between radio-loud and radio-quiet AGN.

\end{document}